\documentclass[runningheads]{llncs}
\usepackage{graphicx}
\usepackage{amssymb}
\usepackage{amsmath}
\usepackage{url}
\usepackage{graphics}
\usepackage{graphicx}
\usepackage{float}
\usepackage{listings}

\begin{document}
\title{A Constraint Model for the Tree Decomposition of a Graph}
\titlerunning{A Constraint Model for Tree Decomposition}
%
\author{Benjamin Bumpus\inst{1}\orcidID{0000-0002-8686-2319} \and
Patrick Prosser\inst{1}\orcidID{0000-0003-4460-6912} \and
James Trimble\inst{1}\orcidID{0000-0001-7282-8745}}
\authorrunning{B. Bumpus et al.}
%
\institute{School of Computing Science, University of Glasgow, Scotland \\
\email{patrick.prosser@glasgow.ac.uk}\\
\url{https://www.gla.ac.uk/schools/computing/}}
\maketitle              
\begin{abstract}
  We present a constraint model for the problem of producing a tree decomposition of a graph.
  The inputs to the model are a simple graph G, the number of nodes in the desired tree
  decomposition and the maximum cardinality of each node in that decomposition. Via a sequence of decision problems, the model allows us to find the
  tree width of a graph whilst delivering a tree decomposition of that width, i.e. a witness. 

\end{abstract}

\section{Introduction}
A tree decomposition of a graph is a mapping from vertices in a graph to nodes in a tree, where the tree nodes are subsets of the
vertices of the graph. The purpose of this is to produce a tree-like strucure of the graph so that, whatever problem that graph
is representing, it can be solved node by node, with a complexity bounded by some function of the size of the nodes. Therefore it is a way of decomposing a problem. Finding a tree decomposition with minimum 
width has been the Holy Grail of the fixed parameter tractability research community
\cite{Bodlaender93,Bodlaender96,BodlaenderFKKT12} and has a long history in Constraint Programming, most notably due to Rina Dechter \cite{DechterP89,Dechter90,GogateD04}, and more recently by Abseher et al \cite{AbseherMW17}.

Our goal is to present what we believe to be the first Constraint  Programming (CP) model for this problem. Our model takes as input a simple graph G and outputs a tree decomposition of that graph, T, with a specified width ($w$) and a specified number of nodes ($m$). If no such tree exists it returns false.  In the next section we present the definition of tree decomposition and treewidth. We then present our CP model, show it running and conclude.

\section{Tree Decomposition and Treewidth: a definition}
Given a simple graph G = (V,E) it may be decomposed into a tree T, where the tree is composed of
$m$ nodes, such that:

\begin{enumerate}
\item A node $N_{i}$ is an improper subset of the vertices of V.
\item The union of all the nodes in the tree is the set of vertices V.
\item If edge (u,v) is in E then that pair of vertices will exist together in at least one of the nodes of T.
\item If vertex v is in node $N_{i}$ and the same vertex is in node $N_{k}$ then vertex v must exist in all
  nodes in the path from $N_{i}$ to $N_{k}$. This can also be expressed as:
for any three nodes $N_{i}$, $N_{j}$ and $N_{k}$,  if $N_{j}$ lies on the path from $N_{i}$ to $N_{k}$ then $N_{i} \cap N_{k} \subseteq N_{j}.$
A third interpretation of this property is that for any vertex $v$ in V, the set of nodes that contain $v$ induces a single subtree of T.
\end{enumerate}

\noindent
Property (4) is sometimes refered to as the \emph{running intersection property}. Conventionally, nodes of T are said to contain \emph{bags}
where a bag is a set of vertices.  To avoid confusion, due to a bag being a synonym for a multi-set, we will say that a node is a set of vertices.
\\ \\
The width of a tree decompostion is the size of the largest node in that decomposition. The treewidth tw(G) is then the minimum width
among all possible tree decompositions of G. Conventionally, treewidth is the minimum width minus one. This is because a tree decomposition of a 
graph G that is itself a tree will have $n-1$ nodes in its tree decomposition T, where each node of T contains a single edge in G, and convention dictates that its treewidth shall be deemed to be one.

\section{The Constraint Model}
We assume that we are given a simple graph G with vertex set V and edge set E. There are $n$ vertices in V and the tree decomposition has exactly $m$ nodes, and each node has cardinality no greater than $w$. We start by giving the constrained variables of the model, then we give the constraints.

\subsection{The Variables}

\begin{flalign}
  & \forall_{i \in [0..m-1]} ~ N_{i} \subseteq \{0, \ldots,n-1\} \label{var1} \\
  & \forall_{i \in [0..m-1]} ~ parent_{i} \in \{0, \ldots, m-1\} \label{var2} \\
  & \forall_{i \in [0..m-1]} ~ depth_{i} \in \{0, \ldots, m-1\} \label{var3} \\
  & \forall_{(u,v) \in E,k \in [0..m-1]} ~ location_{u,v,k} \in \{0,1\} \label{var4} \\
  & \forall_{i,j \in [0..m-1], i<j} ~ intersection_{i,j} \subseteq \{0, \ldots, n-1\} \label{var5} 
\end{flalign}
Constrained set variable $N_{i}$ ((\ref{var1}) above) is a node in the tree decomposition of G, where we are allowed $m$ tree nodes and each node is an improper subset of the  $n$ vertices V (numbered 0 to $n-1$).
\\ \\
The tree decomposition is a rooted tree, where constrained integer variable $parent_{i}$ ((\ref{var2}) above) points to the parent of $N_{i}$, and node $N_{i}$ is at $depth_{i}$
in the tree (constrained integer variable (\ref{var3}) above).  Note that there is no requirement that the tree decomposition be a rooted tree, but making this assumption simplifies our model without loss of generality.
\\ \\
Property (3) insists that each edge appears in at least one node. The constrained integer variable $location_{u,v,k}$ ((\ref{var4}) above) is equal to one
if and only if edge (u,v) is contained in node $N_{k}$. The variable $location_{u,v,k}$ only exists if edge (u,v) exists in E.
\\ \\
We need to maintain the intersections between all pairs of nodes, in order to realise the running intersection property, consequently
constrained set variable $intersection_{i,j}$ ((\ref{var4}) above) is $N_{i} \cap N_{j}$.

\subsection{The Constraints}
Our first constraint, constraint \ref{cons0}, resticts the width of the tree decomposition, such that all nodes have cardinality of at most $w$.

\begin{flalign}
  & \forall_{i \in [0..m-1]} ~ |N_{i}| \leq w \label{cons0}
\end{flalign}

\noindent
Constraint \ref{prop2} ensure that property (2) is respected, i.e. that all vertices in V appear in T.

\begin{flalign}
  & \bigcup_{i =0}^{m-1} ~ N_{i} = V \label{prop2}
\end{flalign}

\noindent
We now capture the intersections between all pairs of nodes, in constraint \ref{cons1}. Note that this is a sparse $m \times m$ array, where 
the constrained set variable $intersection_{i,j}$ is copied into array element $intersection_{j,i}$.

\begin{flalign}
  & \forall_{i,j \in [0..m-1], i<j} ~ intersection _{j,i} \equiv intersection_{i,j} = N_{i} \cap N_{j} \label{cons1} 
\end{flalign}

\noindent
Constraints \ref{tree1} to \ref{tree4} maintain the rooted tree property.  Constraints \ref{tree1} and \ref{tree2} state that node $N_{0}$ is the root of the
tree and is at depth zero. Constraint \ref{tree3} states that all other nodes, $N_{1}$ to $N_{m-1}$, cannot have themselves as parents.
Constraint \ref{tree4} states that if node $N_{j}$ is the parent of $N_{i}$ then the depth of $N_{i}$ is one more than the depth of its parent $N_{j}$.
Constraints \ref{tree1} to \ref{tree4} suffice to ensure that we have a rooted tree.

\begin{flalign}
  & parent_{0}  = 0 \label{tree1} \\
  & depth_{0} = 0 \label{tree2} \\
  & \forall_{i \in [1..m-1]} ~ parent_{i} \neq i \label{tree3} \\
  & \forall_{i \in [1..m-1],j \in [0..m-1], i \neq j} ~ parent_{i} = j \implies depth_{i} = depth_{j} + 1 \label{tree4}
\end{flalign}

\noindent
Constraints \ref{prop3-1} to \ref{prop3-3} ensure that property (3) holds, i.e. that every edge is contained in at least one node
of the tree.  Constraint \ref{prop3-1} allows us to use only the top half of the matrix; because edges are undirected
$location_{u,v,k}$ is exactly the same variable as $location_{v,u,k}$. Constraint \ref{prop3-2} states that $location_{u,v,k}$
takes the value one if and only if the edge (u,v) is contained in the $k^{th}$ node of the tree, and constraint \ref{prop3-3}
insists that the edge (u,v) exists in at least one node.

\begin{flalign}
  & \forall_{(u,v) \in E,k \in [0..m-1]} ~ location_{v,u,k} \equiv location_{u,v,k} \label{prop3-1} \\
  & \forall_{(u,v) \in E,k \in [0..m-1]} ~ location_{u,v,k} = 1 \iff \{u,v\} \subseteq N_{k} \label{prop3-2} \\
  & \forall_{(u,v) \in E} ~ \sum_{k=0}^{m-1} location_{u,v,k} \geq 1 \label{prop3-3} 
\end{flalign}

\noindent
Finally we have constraint \ref{prop4}, to enforce the running intersection property (4). If we have two distinct nodes $N_{i}$ and $N_{k}$ that are at the 
same  depth, or node $N_{k}$ is deeper in the tree than $N_{i}$, then if we relax the property that the tree is rooted, then the parent of $N_{k}$ is on the path from $N_{i}$ to $N_{k}$, and we insist that every vertex that is common to $N_{i}$ and $N_{k}$, i.e. $intersection_{i,k}$ is subsumed by the vertices in the parent node of $N_{k}$, namely $N_{parent_k}$. 

\begin{flalign}
  & \forall_{i,k \in [0..m-1], i \neq j} ~ depth_{i} \leq depth_{k} \implies intersection_{i,k} \subseteq N_{parent_{k}} \label{prop4} 
\end{flalign}

\section{Implementation}
We have endeavoured to use only constraints that we should expect to see in any
Constraint Programming toolkit. This is one of the reasons why we did not use the tree constraint
(such as \cite{prosser2006}).
In implementing our model, constraint \ref{prop4} required the use of the element
constraint, where a constrained integer variable is used as an index into an array of constrained set variables.
This is the most sophisticated constraint in our model.

The model was implemented in the choco4 CP toolkit \cite{choco4}. The decision variables were the
$parent$ variables combined with the flattened $location$ variables. This was a convenience so that we could use
library variable and value ordering heuristics, rather than have something more complex that allowed
us to mix constrained integer variables with constrained set variables. 

Simple symmetry breaking was added. The nodes were channelled to bits sets, such that bit set $B_{i,j} = 1 \iff j \in N_{i}$. 
A lexicographical ordering was then posted between bit sets such that $B_{i} \preceq B_{i+1}$.

\section{Computational Experience}
Finding the tree width of a graph is done as a sequence of decision problems. We can start with
the number of nodes $m$ in the tree to be equal to 1, and the width $w = n$.  A single node tree is then found trivially. 
We then increment $m$ and decrement $w$ and repeat the process until $w = 2$ (in which case G is a tree) or failure is reported, and the 
previous values of $m$ and $w$ give us an optimal tree decomposition of minimum width $w-1$. This process is sound and complete:

\begin{proof}
Proposition: If $T$ is a tree decomposition of $m$ nodes and width $w$ of graph $G$, with no duplicate nodes, then $m \leq n - w + 1$,
where $n$ is the number of vertices in $G$.  By way of contradiction,  suppose otherwise, then all tree decompositions of $G$ would require a tree with more than $n - w + 1$ nodes. But since we do not allow duplicate nodes, every node must contain one ``unseen'' vertex, hence 
$n \geq w + m - 1 > w + (n - w + 1) - 1 = n$, and that is a contradiction. $\square$
\end{proof}

\begin{figure}[h]
\includegraphics[height=8.6cm,width=11.0cm]{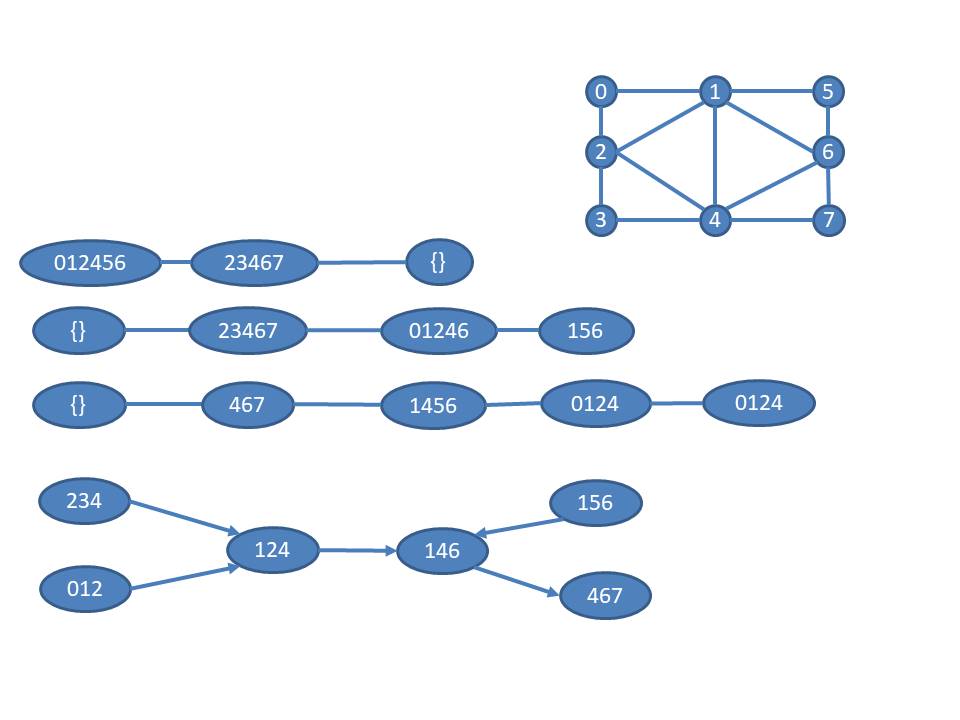}
\caption{Input graph in top right quandrant. Four tree decompositions. From top to bottom, m=3 and w = 6, m = 4 and w = 5, m = 5 and w = 4, m = 6 and w = 3. Last tree decomposition, directed $parent$ pointers are shown.}
\label{fig1}
\end{figure}

In Figure \ref{fig1} we show the results of this process on a graph with 8 vertices and 13 edges. We show the tree decompositions 
for $m = 3$ and $w = 6$, and onwards to the optimal decomposition $m = 6$ and $w = 3$.  For $m = 5$ and $w = 4$ our model required 2.4 seconds of resolution time and 7,388 decisions.  For $m = 6$ and $w = 3$ our model required 1.2 seconds of resolution time and 2,959 decisions.  Optimality, 
the unsatisfiability of $m = 7$ and $w = 2$, is proved in 0.2 seconds and 197 decisions.

We do not include detailed experimental comparisons with existing algorithms, but we observe that our model is orders of magnitude slower than state-of-the-art methods for tree decomposition. Rather than being of practical use as a solver, we present our model as a way of exploring and explaining tree width to an interested user, and as a tool that could easily be extended to handle side constraints. It might be likened to an easily assembled and easily adapted Tinkertoy.

\section{Conclusion}
We present what we believe to be the first CP model for tree decomposition of a graph, with a process that allows us to determine the tree width of a graph. The model is simple, and we believe it can be readily encoded in most any CP modelling languages. Our model, in its current form, is too slow
to be of pratical use, but it might have a place as a dynamic tool to help those who ``want to get to know about tree width." Our model is also adaptable. For example, we can find the path width of a graph via a simple edit: $\forall_{0 < i < m} ~ parent_{i} = i-1$.

\end{document}